\documentclass[final]{raa_rb}
\usepackage{times}             
\usepackage{natbib}
\usepackage{amssymb,amsmath}
\bibpunct{(}{)}{;}{a}{}{,}
\usepackage{graphicx}
\usepackage{xcolor}
\usepackage{multirow}
\usepackage[T1]{fontenc}
\usepackage{newtxtext, newtxmath}
\usepackage{float}
\usepackage{longtable}
\usepackage{enumitem}
\usepackage{longtable,listings}
\usepackage[flushleft]{threeparttable}
\usepackage{parcolumns, hyperref}
\def\oc3{[O~{\sc iii}]$_c$}
\def\ob3{[O~{\sc iii}]$_b$}

\def\oc{[O~{\sc iii}]$_c$}
\def\ob{[O~{\sc iii}]$_B$}

\def\d4{D{\it n}4000}

\begin{document}

\title{Dependence of 4000\AA~ break strength on narrow emission line properties in local Type-2 AGN}

\author{Xue-Guang Zhang
\inst{1}
}

\institute{
Guangxi Key Laboratory for Relativistic Astrophysics, School of Physical Science and Technology,
GuangXi University, Nanning, 530004, P. R. China}

\date{}

\abstract{
	In this manuscript, we report evidence to support the dependence of D{\it n}4000 (4000\AA~ break strength to 
trace stellar ages) on central AGN activity traced by narrow emission line properties in local Type-2 AGN in SDSS DR16. Based 
on the measured D{\it n}4000 and flux ratios of [O~{\sc iii}] to narrow H$\beta$ (O3HB) and [N~{\sc ii}] to narrow 
H$\alpha$ (N2HA) and narrow H$\alpha$ line luminosity $L_{H\alpha}$, linear dependence of the D{\it n}4000 on the O3HB, 
N2HA and $L_{H\alpha}$ in the local Type-2 AGN can provide clues to support the dependence of D{\it n}4000 on properties 
of narrow emission lines. Linear correlations between the D{\it n}4000 and the O3HB and N2HA can be found in the local 
Type-2 AGN, with Spearman rank correlations about -0.39 and 0.53. Meanwhile, stronger dependence of the D{\it n}4000 
on the $L_{H\alpha}$ can be confirmed in Type-2 AGN, with Spearman rank correlation coefficient about -0.7. Moreover, combining 
the $L_{H\alpha}$ and the N2HA, a more robust and stronger linear correlation can be confirmed between the D{\it n}4000 
and the new parameter $LR=0.2\log(L_{H\alpha})-0.5\log(\rm N2HA)$, with Spearman rank correlation coefficient about -0.76 
and with smaller RMS scatters. After considering necessary effects, the dependence of D{\it n}4000 on $LR$ is stable and 
robust enough for the local Type-2 AGN, indicating the $LR$ on the narrow emission lines can be treated as a better indicator to 
statistically trace stellar ages of samples of more luminous AGN with weaker host galaxy absorption features.
\keywords{Galaxies:active - Galaxies:nuclei - Galaxies:emission lines - Galaxies:Seyfert}
}

\authorrunning{Zhang}            
\titlerunning{Dependence of D{\it n}4000 on narrow emission lines}

\maketitle

\section{Introduction}
 
	Tight connections between Active Galactic Nuclei (AGN) and host galaxies can be expected due to AGN feedback through 
galactic-scale outflows playing key roles in galaxy evolutions as discussed in \citet{ref1, ref2, ref3, ref4, ref5, ref6, 
ref7}. Either negative or positive AGN feedback on star-formations in AGN host galaxies have been reported and studied in both 
observational and theoretical results, such as clear evidence on negative AGN feedback in \citet{ref9, ref15, ref11, ref8, 
zh24a} and on positive AGN feedback in \citet{ref13, ref10, ref12}. Due to the contrary conclusions on effects of AGN feedback, 
it is interesting to provide further clues on effects of AGN feedback, through different methods.  

\begin{figure}
	\centering\includegraphics[width=\textwidth]{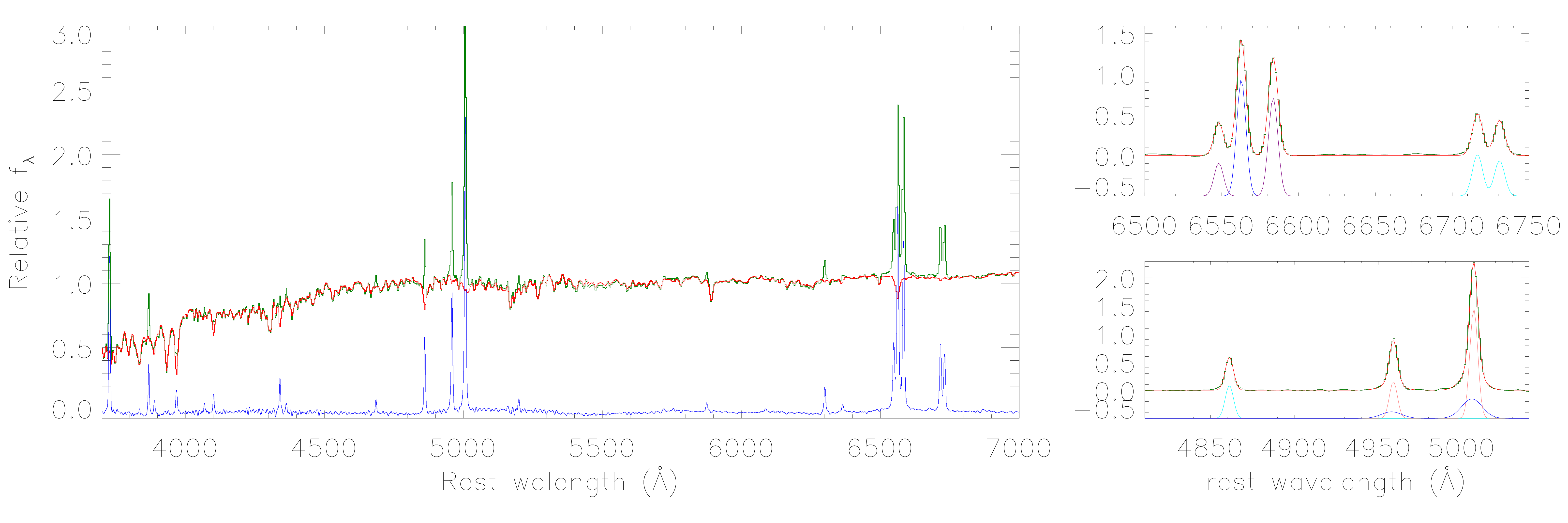}
\caption{Left panel shows the SSP method determined best fitting results (solid red line) to the host galaxy features in the 
inverse variance weighted mean spectrum (solid dark green line) of the Type-2 AGN, and the corresponding line spectrum (solid 
blue line) after subtractions of the host galaxy contributions. Right panels show the best fitting results (solid red line) to 
the emissions lines (solid dark green line) around H$\alpha$ (top right panel) and around H$\beta$ (bottom right panel) in the 
mean spectra of Type-2 AGN, after subtractions of the host galaxy contributions. In top right panel, solid lines in blue, in 
purple and in cyan show the determined narrow H$\alpha$, [N~{\sc ii}] and [S~{\sc ii}] doublets, respectively. In bottom right 
panel, solid lines in blue, in pink and in cyan show the determined extended and core components of [O~{\sc iii}] doublet, and 
narrow H$\beta$, respectively.}
\label{msp}
\end{figure}

\begin{figure}
	\centering\includegraphics[width = \textwidth]{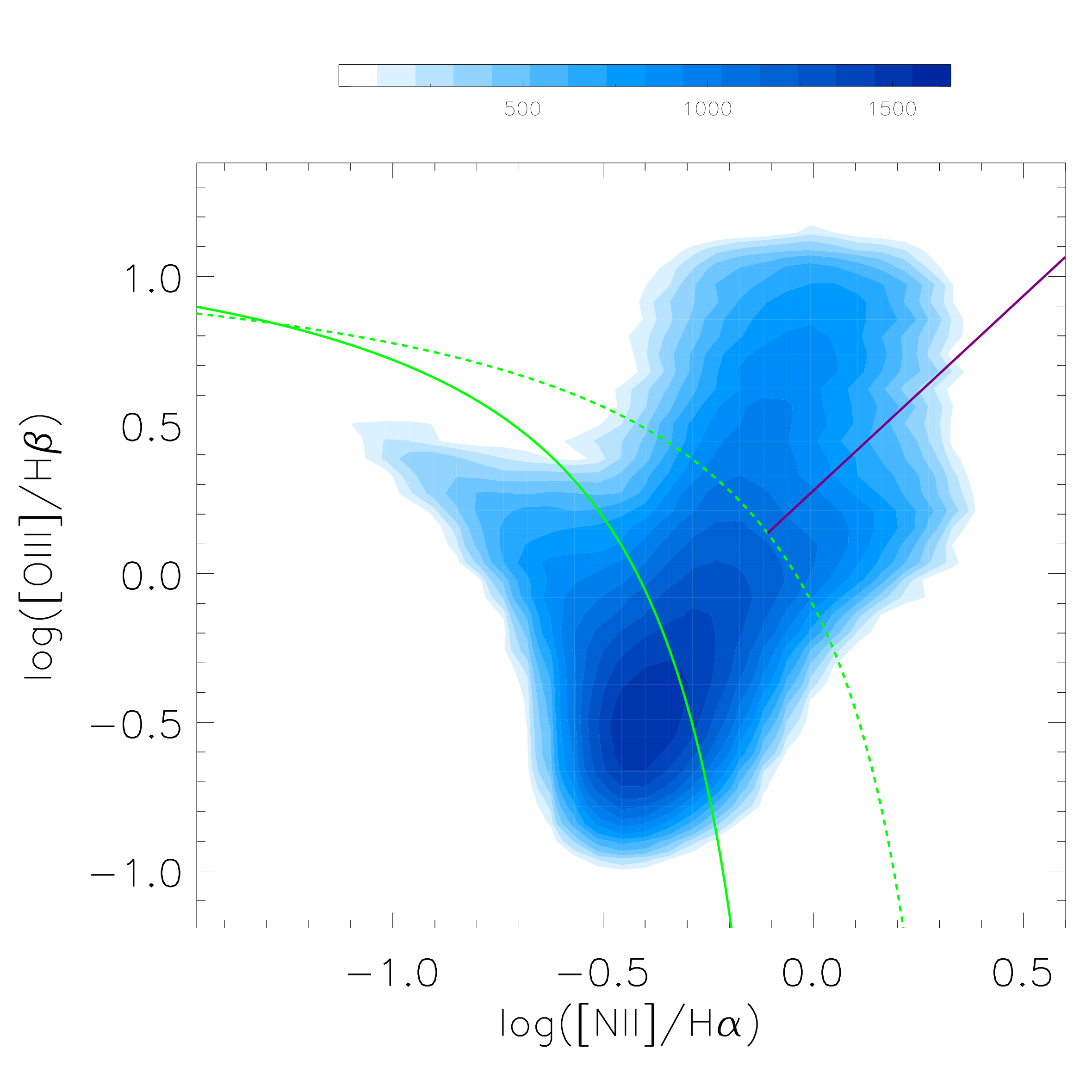}
\caption{The BPT diagram of O3HB versus N2HA for all the 87828 narrow emission line main galaxies in SDSS DR16, with all the 
collected 14031 Type-2 AGN lying above the dividing line shown as dashed green line and all the collected 44501 HII galaxies 
lying below the dividing line shown as solid green line. Solid purple line shows the dividing line between Seyfert-2 galaxies 
and LINERs in this manuscript. The color bar in top corner represents the corresponding number densities.}
\label{bpt}
\end{figure}

	Besides direct measurements of star formations, AGN feedback should have apparent and strong effects on stellar ages 
in AGN host galaxies. Different methods have been proposed to determine the mean stellar ages of galaxies. \citet{cf05} have 
shown the measurements of light-weighted mean stellar ages by the STARLIGHT code based on the SSP (Simple Stellar Populations) 
methods \citep{ref33}. \citet{ref32} have shown that the Penalized Pixel-Fitting method can lead to reliable and smoother 
star-formation histories after considering the regularization method, then lead to estimated mass-weighted stellar ages. 
\citet{ka03a} have estimated mean stellar ages by the parameters of D{\it n}4000 (4000\AA~ break strength) being applied 
to trace starforming histories, similar discussions on D{\it n}4000 as an indicator to trace stellar ages can be found 
in \citet{zd15}. More recently, \citet{gs20} have shown the loose dependence of D{\it n}4000 on [O~{\sc iii}] line 
luminosity, indicating narrow line luminosity could be roughly applied to trace stellar age properties.

	Therefore, there are reliable methods to estimate stellar ages in AGN host galaxies. Certainly, in order to clearly 
estimate stellar ages, host galaxy features should be apparently significant in spectra, indicating Type-2 AGN (narrow 
emission line AGN) rather than Type-1 AGN (broad line AGN) are the better candidates to study effects of AGN activity on stellar 
ages. Moreover, based on the commonly accepted unified model \citep{ref25, ref26} of AGN, emissions from central accretion 
disks and from central broad emission line regions are totally obscured by central dust torus in Type-2 AGN, indicating that 
both spectroscopic continuum emissions and absorption features of Type-2 AGN can be clearly applied to measure the mean stellar 
ages with few contaminations.

	Besides the host galaxy features in Type-2 AGN to determine stellar ages traced by the D{\it n}4000 parameter mainly 
considered in this manuscript, AGN activity of Type-2 AGN can be well traced by narrow emission line properties in the well-known 
BPT diagrams \citep{bpt, kb01, ref21, kb06, ref20, ref22}. Therefore, there are sufficient conditions to do the research on potential 
dependence of stellar ages on central AGN activity in a large sample of Type-2 AGN which can provide clear clues on effects of AGN 
feedback. This manuscript is organized as follows. Section 2 presents the data samples of Type-2 AGN, and the simple procedure to 
measure the D{\it n}4000 parameter. Section 3 presents our main results and discussions. Section 4 gives our final summary 
and conclusions. And in this manuscript, we have adopted the cosmological parameters of $H_{0}=70{\rm km\cdot s}^{-1}{\rm Mpc}^{-1}$, 
$\Omega_{\Lambda}=0.7$ and $\Omega_{\rm m}=0.3$.

\section{Data Samples of Type-2 AGN} 

	All the 87828 low redshift narrow emission line galaxies are firstly collected from the main galaxies in SDSS DR16 (Sloan 
Digital Sky Survey, Data Release 16, \citealt{ref31}) as the parent sample, with redshift smaller than 0.35 and with median spectral 
signal-to-noise ratio ($S/N$) larger than 10 and with SDSS provided reliable narrow emission lines of H$\beta$, [O~{\sc iii}], 
H$\alpha$ and [N~{\sc ii}] and with reliable SDSS pipeline measured stellar velocity dispersions and with Balmer decrement (flux 
ratio of narrow H$\alpha$ to narrow H$\beta$) smaller than 6. Meanwhile, we should note that the SDSS pipeline provided emission 
line parameters have been collected from the database of 'GalSpecLine' reported by MPA-JHU as described in \citet{br04, tr04, ka03a}, 
and the D{\it n}4000 are collected from the public SDSS database of 'GalSpecIndx'. Moreover, besides the SDSS pipeline provided 
values, the parameters of emission lines and D{\it n}4000 have also been re-measured as follows.


	Narrow emission lines have been re-measured in the collected narrow emission line objects, after subtractions of host 
galaxy contributions, in order to confirm the reliability of the narrow emission line parameters. Similar as what we have recently 
done in \citet{ref37, zh21b, zh21c, zh22b, zh22c, zh23, zh24a}, two main steps are applied. The first step is to determine the host 
galaxy contributions by the SSP method, as discussed in detail in \citet{ref33, ref21, cf05, ref32}. The second step is to measure 
the emission lines by multiple Gaussian functions in the line spectrum after subtractions of host galaxy contributions. Here, we 
do not show further detailed discussions on the SSP method nor detailed discussions on measurements of emission lines any more, 
but Fig.~\ref{msp} shows an example on the SSP method determined host galaxy contributions in the inverse variance weighted mean 
spectrum of the collected Type-2 AGN, and the best fitting results to the emission lines, through the Levenberg-Marquardt least-squares 
minimization technique. Here, we do not show further discussions on the measured line parameters, but there are well consistent 
line parameters between our measured values and the reported values in 'GalSpecLine' for the narrow emission line galaxies, the 
correlations between SDSS provided line intensities and our measured line intensities having Spearman Rank correlation coefficients 
larger than 0.93, to support the reliability of our measured emission line parameters.

	Furthermore, besides the database 'GalSpecIndx' provided parameter D{\it n}4000 (parameter d4000\_n in the database), 
D{\it n}4000 are re-measured through the downloaded SDSS spectra. The 4000\AA~ break strength $Dn4000=\frac{Con_R}{Con_B}$ 
can be measured by the definition described in \citet{ba99, ka03a}, with $Con_B$ and $Con_R$ as mean continuum emission intensities 
within rest wavelength from 3850\AA~ to 3950\AA~ and from 4000\AA~ to 4100\AA. Here, we do not show further discussions on our 
measured D{\it n}4000, but there are well consistent D{\it n}4000 between our measured values and the reported values 
in 'GalSpecIndx' for the narrow emission line galaxies, with Spearman Rank correlation coefficient larger than 0.9, to support 
the reliability of our measured D{\it n}4000.

	Based on the measured narrow line emission flux ratios of [O~{\sc iii}]$\lambda5007$\AA~ to narrow H$\beta$ (O3HB) and 
[N~{\sc ii}]$\lambda6583$\AA~ to narrow H$\alpha$ (N2HA) applied in the BPT diagram, the dividing line in \citet{kb01}
\begin{equation}
	\log(\rm O3HB)~=~\frac{0.61}{\log(\rm N2HA)~-~0.47}~+~1.19
\end{equation}
is accepted to collect 14031 Type-2 AGN and to exclude the composite objects in our collected Type-2 AGN. The mean spectrum of the 
14031 Type-2 AGN is shown in left panel of Fig.~\ref{msp}. Meanwhile, the dividing line in \citet{ref21}
\begin{equation}
	\log(\rm O3HB)~=~\frac{0.61}{\log(\rm N2HA)~-~0.05}~+~1.30
\end{equation}
is accepted to collect 44501 HII galaxies and to exclude the composite objects in our collected HII galaxies. The BPT diagram of 
O3HB versus N2HA and the applied dividing lines are shown in Fig.~\ref{bpt}.

	Based on the measured D{\it n}4000 and the line parameters of narrow emission lines, it is interesting to check 
dependence of the D{\it n}4000 on the narrow emission line properties. 

\section{Main results and Discussions}

\begin{figure}
	\centering\includegraphics[width = \textwidth]{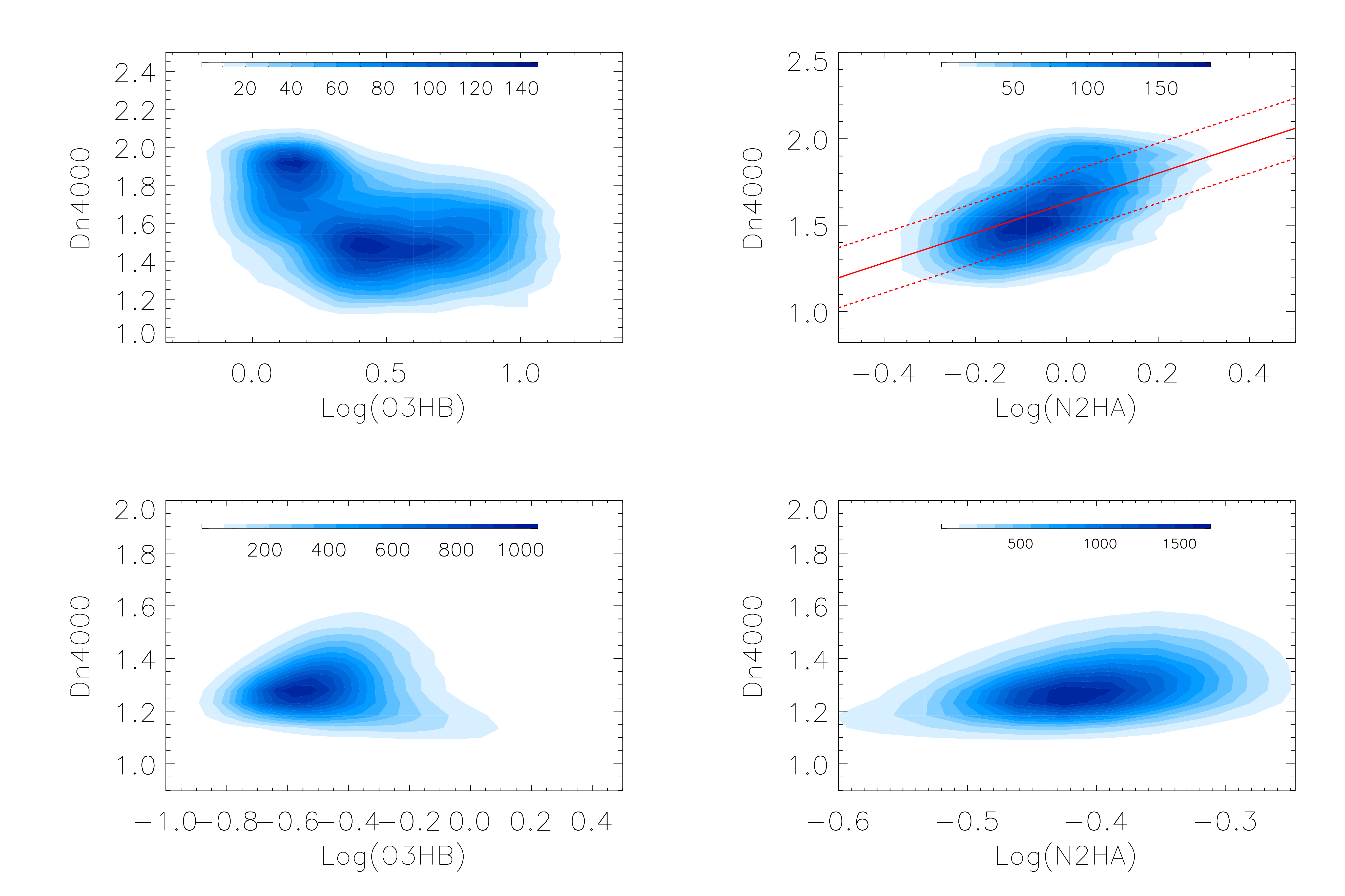}
\caption{On the correlations between D{\it n}4000 and O3HB (left panels) and between D{\it n}4000 and N2HA 
(right panels) in the collected Type-2 AGN (top panels) and HII galaxies (bottom panels). In top right panel, solid red line and 
dashed red lines show the best fitting results and the corresponding 1RMS scatters.}
\label{res}
\end{figure}

	For the collected 14031 Type-2 AGN, correlations between D{\it n}4000 and $\log(\rm N2HA)$ and between 
D{\it n}4000 and $\log(\rm O3HB)$ are shown in top panels of Fig.~\ref{res}. The Spearman Rank correlation 
coefficients are about -0.39 ($P_{null}<10^{-15}$) and 0.53 ($P_{null}<10^{-15}$) for the D{\it n}4000 versus O3HB and for the 
D{\it n}4000 versus N2HA, respectively, indicating the D{\it n}4000 more sensitively depends on the N2HA than on the 
O3HB. For the stronger correlation between D{\it n}4000 and $\log(\rm N2HA)$ in Type-2 AGN, after considering the 
uncertainties in both coordinates, the corresponding best fitting results can be determined by the commonly applied Least Trimmed 
Squares (LTS) robust technique \citep{ref23, ref24} 
\begin{equation}
	D{\it n}4000~=~(1.628\pm0.002)~+~(0.864\pm0.011)\log(\rm N2HA)
\end{equation}
with the RMS scatter about 0.174. Meanwhile, considering larger O3HB leading to stronger central AGN activity, Type-2 AGN with 
stronger AGN activity (larger O3HB) should have younger stellar ages, which are consistent with the conclusions in \citet{ref21} 
that the host galaxies of high-luminosity AGN have much younger mean stellar ages.

	Moreover, the dependence of D{\it n}4000 on O3HB and on N2HA are also checked in the collected 44501 HII galaxies 
and shown in bottom panels of Fig.~\ref{res}. The Spearman Rank correlation coefficients are about 0.01 ($P_{null}<10^{-15}$) and 
0.37 ($P_{null}<10^{-15}$) for the D{\it n}4000 versus O3HB and for the D{\it n}4000 versus N2HA in the HII galaxies, 
respectively, indicating weak dependence of D{\it n}4000 on the narrow line flux ratios in HII galaxies. Furthermore, as 
the results shown in Fig.~\ref{res}, there are similar spans of the applied N2HA in the Type-2 AGN with 
$\log(\rm N2HA)\in[\sim-0.3,~\sim0.3]$ and in the HII galaxies with $\log(\rm N2HA)\in[\sim-0.55,~\sim-0.25]$, 
therefore, the different dependence of D{\it n}4000 on N2HA between Type-2 AGN and HII galaxies are not due to different 
spans of the applied N2HA. Actually, the span of $\log(\rm N2HA)$ is slightly larger in the Type-2 AGN, if there were 
effects of the spans, more loose dependence of D{\it n}4000 on N2HA should be expected in the Type-2 AGN. 

\begin{figure*}
	\centering\includegraphics[width = \textwidth]{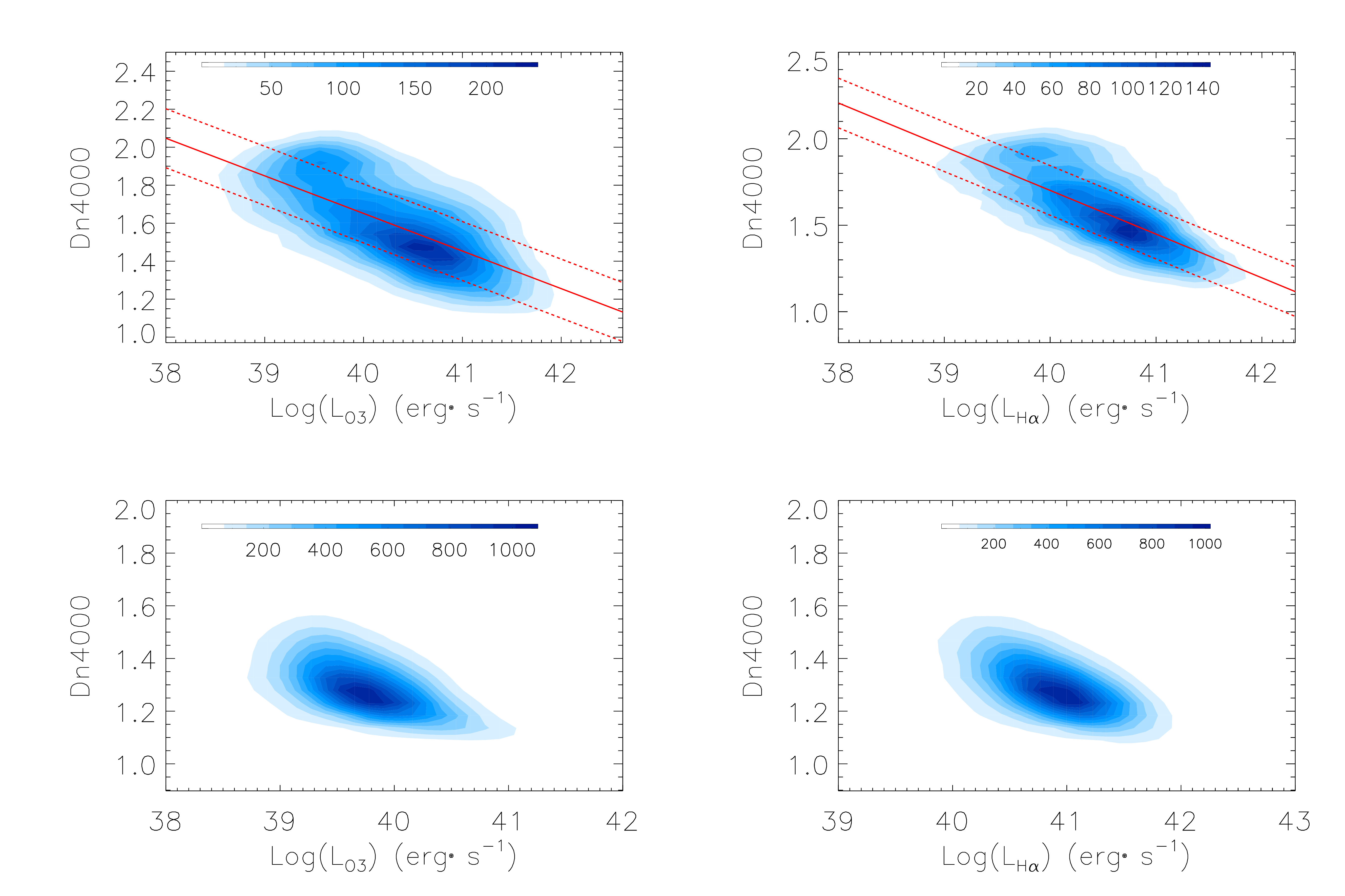}
\caption{On the correlations between D{\it n}4000 and $L_{O3}$ (left panels) and between D{\it n}4000 and 
$L_{H\alpha}$ (right panels) in the collected Type-2 AGN (top panels) and in the HII galaxies (bottom panels). In each top panel, 
solid red line and dashed red lines show the best fitting results and the corresponding 1RMS scatters.}
\label{res2}
\end{figure*}

	Besides the dependence of D{\it n}4000 on O3HB and N2HA applied in the known BPT diagrams, dependence of 
D{\it n}4000 on narrow line luminosities are also checked in Type-2 AGN and in HII galaxies, and shown in Fig.~\ref{res2}. 
The correlations between D{\it n}4000 and [O~{\sc iii}]$\lambda5007$\AA~ line luminosity ($L_{O3}$) have Spearman rank 
correlation coefficients about -0.66 ($P_{null}<10^{-15}$) and -0.53 ($P_{null}<10^{-15}$) in Type-2 AGN and in HII galaxies, 
respectively. And the correlations between D{\it n}4000 and narrow H$\alpha$ line luminosity ($L_{H\alpha}$) have 
Spearman rank correlation coefficients about -0.70 ($P_{null}<10^{-15}$) and -0.55 ($P_{null}<10^{-15}$) in Type-2 AGN and in HII 
galaxies, respectively. And for the stronger correlations between D{\it n}4000 and $L_{H\alpha}$ and between 
D{\it n}4000 and $L_{O3}$ in Type-2 AGN, with considering the uncertainties in both coordinates, the corresponding 
best-fitting results can be determined by the commonly applied the LTS robust technique
\begin{equation}
\begin{split}
	&D{\it n}4000=(11.811\pm0.085)-(0.253\pm0.002)\log(\frac{L_{H\alpha}}{\rm erg\cdot~s^{-1}}) \\
	&D{\it n}4000=(9.569\pm0.075)-(0.198\pm0.002)\log(\frac{L_{O3}}{\rm erg\cdot~s^{-1}})
\end{split}
\end{equation}
with the RMS scatters about 0.143 and 0.155, respectively.

\begin{figure*}
	\centering\includegraphics[width = \textwidth]{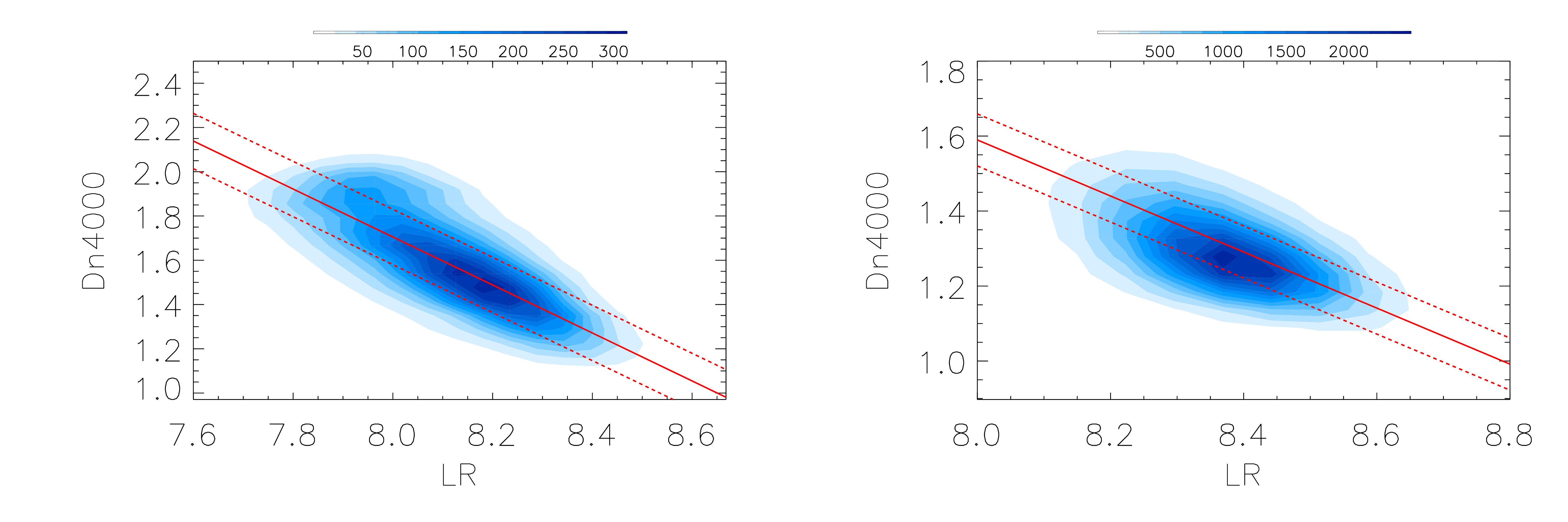}
\caption{On the correlations between D{\it n}4000 and $LR=0.2\log(L_{H\alpha})-0.5\log(\rm N2HA)$ in the Type-2 
AGN (left panel) and in HII galaxies (right panel). Solid red line, dotted red lines show the best fitting results to the 
correlation and the corresponding 1RMS scatters, respectively.}
\label{d4}
\end{figure*}

	In Type-2 AGN, the dependence of D{\it n}4000 on N2HA and on narrow H$\alpha$ line luminosity are apparent with linear 
correlation coefficients of 0.53 and -0.70, but not strong enough. Probably, combining narrow emission line flux ratios and 
the emission line luminosities could lead to a stronger linear correlation on the D{\it n}4000 parameter. Here, a new 
parameter $LR$ is defined as
\begin{equation}
	LR~=~0.2~\times~\log(L_{H\alpha}/{\rm erg\cdot~s^{-1}})~-~0.5~\times~\log(\rm N2HA)
\end{equation}  
Left panel of Fig.~\ref{d4} shows the correlation between D{\it n}4000 and the new parameter $LR$ in the Type-2 AGN, with 
the Spearman rank correlation coefficient about -0.76 ($P_{null}<10^{-15}$). Meanwhile, right panel of Fig.~\ref{d4} shows the 
correlation between the D{\it n}4000 and the new parameter $LR$ in the HII galaxies, with the Spearman rank correlation 
coefficient about -0.65 ($P_{null}<10^{-15}$). The correlations between D{\it n}4000 and $LR$ in Type-2 AGN and in HII 
galaxies can be determined by the LTS robust technique
\begin{equation}
\begin{split}
	&D{\it n}4000=(10.377\pm0.056)-(1.084\pm0.007)LR \ \ (AGN) \\
	&D{\it n}4000=(7.569\pm0.029)-(0.747\pm0.003)LR \ \ (HII)
\end{split}
\end{equation}
with the RMS scatters about 0.125 and 0.069, respectively. Here, simple descriptions are given on the determined formula of 
$LR$. Accepted the formula $LR~=~A~\times~\log(L_{H\alpha}/{\rm erg\cdot~s^{-1}})~+~B~\times~\log(\rm N2HA)$ with $A$ and $B$ 
randomly collected values from -10 to 10, after checking the corresponding Spearman Rank correlation coefficients $SR$ for the 
correlation between $LR$ and $D{\it n}4000$, the maximum $SR$ leads to the accepted $A=0.2$ and $B=-0.5$. Moreover, among 
all the emission line parameters, $LR$ of combining H$\alpha$ luminosity ($0.2\log(L_{H\alpha}$) and N2HA ($-0.5\log(\rm N2HA)$) 
leads to the strongest linear correlation between D{\it n}4000 and $LR$. And there are no further discussions of $LR$ combining 
with different other narrow emission line parameters.

\begin{figure*}
	\centering\includegraphics[width = \textwidth]{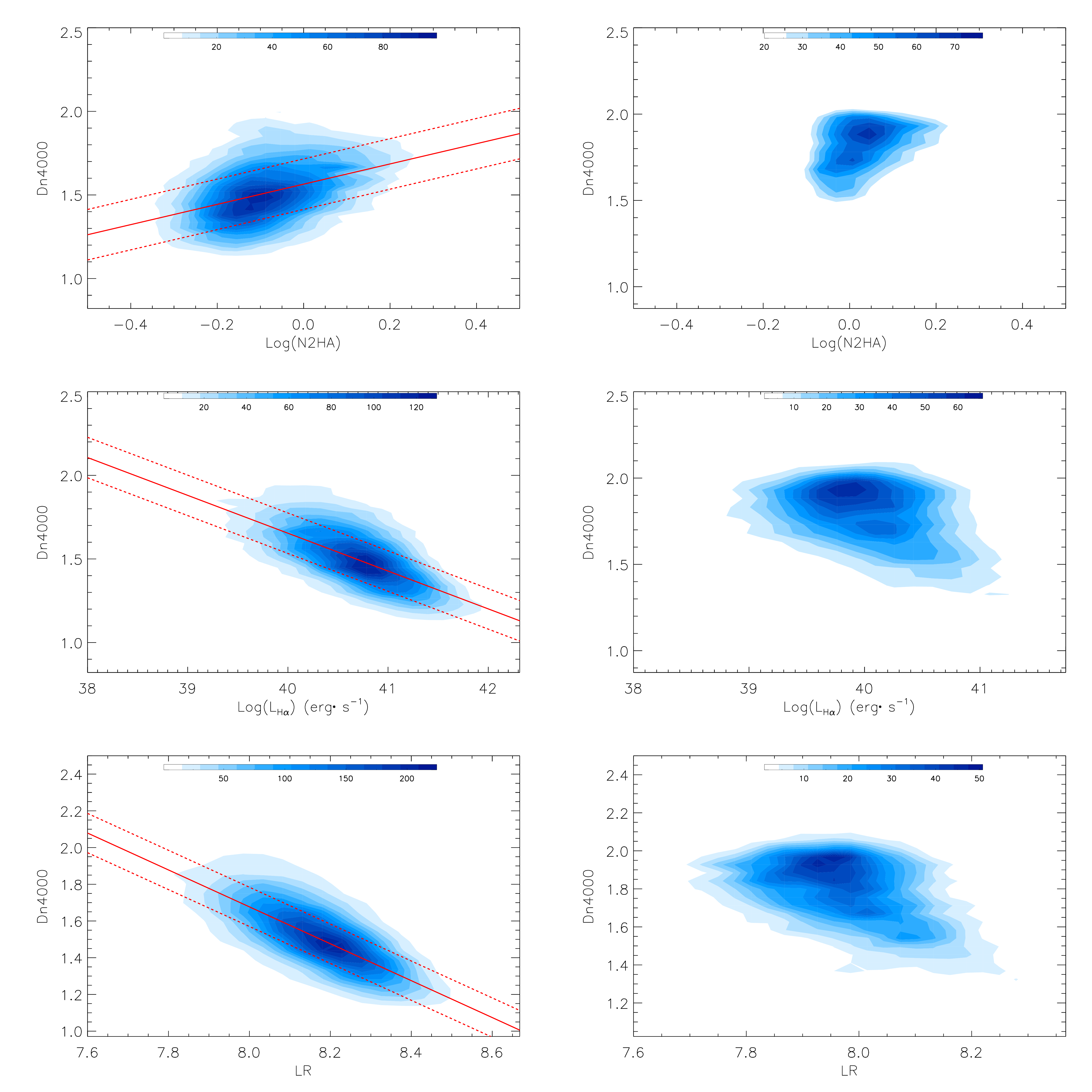}
\caption{On the correlations between D{\it n}4000 and N2HA (top panels) and between D{\it n}4000 and $L_{H\alpha}$ 
(middle panels) and between D{\it n}4000 and $LR$ (bottom panels) in the Seyfert-2 galaxies (left panels) and in LINERs 
(right panels). In each left panel, solid red line, dotted red lines show the best fitting results to the correlation and the 
corresponding 1RMS scatters, respectively.}
\label{refL}
\end{figure*}

	Besides the results shown and discussed above, LINERs (Low Ionization Nuclear Emission Line Regions) \citep{ht80, ft92, 
ds96, mm17} are included in the sample of Type-2 AGN. LINERs and commonly classified Seyfert-2 galaxies lie in two apparently 
different sub-branches in the BPT diagrams. Therefore, it is necessary and interesting to check whether LINERs have apparent 
effects on the results on D{\it n}4000. Here, we do not discuss whether LINERs are genuine AGN. How to classify LINERs in 
the BPT diagrams have been well discussed in \citet{gk06, kb06, jd11, ca18, as21}, there are no further discussions on classifications 
of LINERs in BPT diagrams. In the BPT diagram of O3HB versus N2HA in Fig.~\ref{bpt}, considering the dividing lines shown 
as dashed green line and solid purple line, there are 4134 classified LINERs lying above the dashed green line and below the 
solid purple line. Here, there are no further discussions on the dividing line between Seyfert-2 galaxies and LINERs in the 
BPT diagram of O3HB versus N2HA. However, if the classifications of LINERs are accepted through the dividing lines reported in 
the BPT diagram of O3HB versus S2HA (flux ratio of [S~{\sc ii}] to narrow H$\alpha$) reported in \citet{kb06}, the corresponding 
locations of LINERs and Seyfert-2 galaxies in the BPT diagram of O3HB versus N2HA can roughly lead to the dividing line shown 
as purple solid line in Fig.~\ref{bpt}.

	Fig.~\ref{refL} shows the corresponding dependence of D{\it n}4000 in the 9897 Seyfert-2 galaxies and in the 4134 LINERs. 
The correlations between D{\it n}4000 and N2HA have Spearman rank correlation coefficients about 0.40 ($P_{null}<10^{-15}$) 
and 0.25 ($P_{null}<10^{-15}$) in the Seyfert-2 galaxies and in the LINERs, respectively. The correlations between D{\it n}4000 
and $L_{H\alpha}$ have Spearman rank correlation coefficients about -0.66 ($P_{null}<10^{-15}$) and -0.37 ($P_{null}<10^{-15}$) in 
the Seyfert-2 galaxies and in the LINERs, respectively. The correlations between D{\it n}4000 and $LR$ have Spearman rank 
correlation coefficients about -0.72 ($P_{null}<10^{-15}$) and -0.42 ($P_{null}<10^{-15}$) in the Seyfert-2 galaxies and in the 
LINERs, respectively. The very different correlations probably indicate LINERs having unique physical properties different from 
Seyfert-2 galaxies. Meanwhile, the linear correlations in the 9897 Seyfert-2 galaxies can be determined by the LTS technique 
\begin{equation}
\begin{split}
	&D{\it n}4000=(1.565\pm0.002)+(0.605\pm0.012)\log(\rm N2HA) \\
	&D{\it n}4000=(10.706\pm0.094)-(0.226\pm0.002)\log(\frac{L_{H\alpha}}{\rm erg\cdot s^{-1}})  \\
	&D{\it n}4000=(9.717\pm0.067)-(1.005\pm0.008)LR
\end{split}
\end{equation}
with the RMS scatters about 0.151, 0.121 and 0.107, respectively. It is clear that LINERs have few effects on the shown results 
on D{\it n}4000 traced by emission line parameters in Type-2 AGN. Meanwhile, the dependence of D{\it n}4000 on 
emission line properties are quite weaker in LINERs than in normal Seyfert-2 galaxies. 

	Before ending the section, seven additional points should be noted as follows. 

\begin{figure*}
	\centering\includegraphics[width = \textwidth]{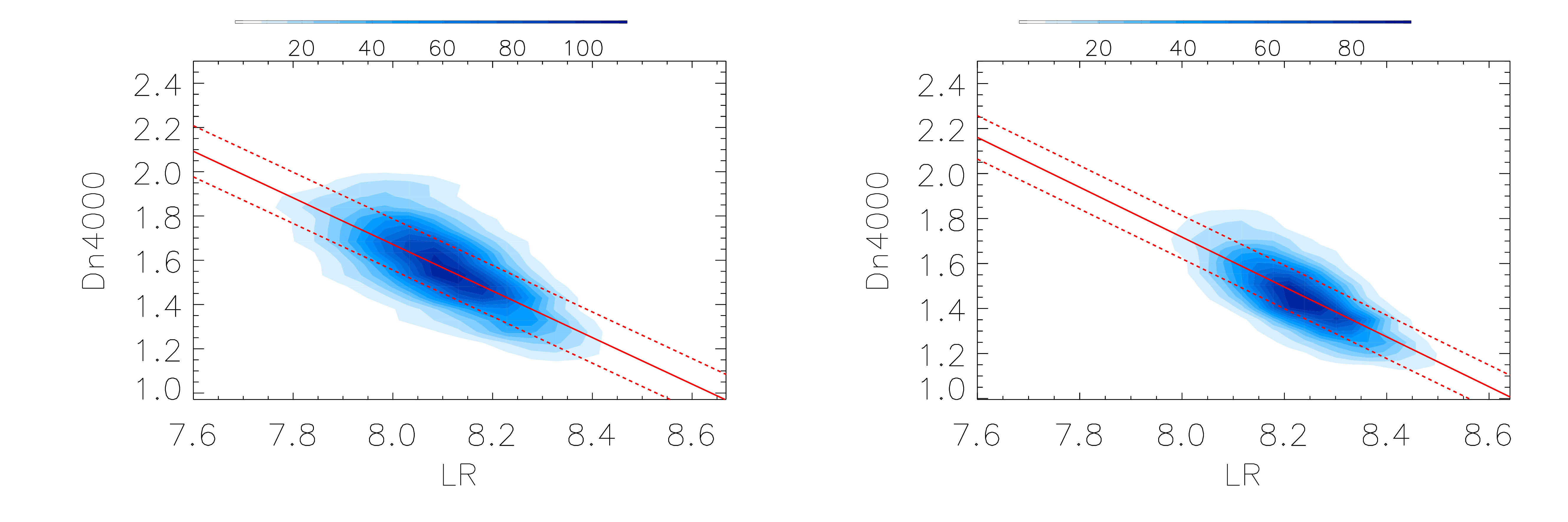}
\caption{On the correlation between D{\it n}4000 and $LR$ in the Seyfert-2 galaxies with redshift smaller than 0.1 (left 
panel) and with redshift larger than 0.1 (right panel). In each panel, solid red line, dotted red lines show the best fitting 
results to the correlation and the corresponding 1RMS scatters, respectively.}
\label{cz}
\end{figure*}

	For the first point, it should be necessary to discussed effects of spectral signal-to-noise ($S/N$) on the final results, 
because only the low redshift narrow emission line Type-2 AGN with $S/N>10$ are considered, as discussed in Section 2. Among the 
collected 9897 Seyfert-2 galaxies (LINERs not considered), correlations between D{\it n}4000 and N2HA and between D{\it n}4000  
and $L_{H\alpha}$ and between D{\it n}4000 and $LR$ are carefully checked for the 3503 Seyfert-2 galaxies with $S/N>20$. The 
correlations have corresponding Spearman rank correlation coefficients about 0.34 ($P_{null}<10^{-15}$), about -0.67 
($P_{null}<10^{-15}$) and about -0.71 ($P_{null}<10^{-15}$) for the Seyfert-2 galaxies with $S/N>20$. Here, the correlations 
are not shown for the 3503 Seyfert-2 galaxies with $S/N>20$, but are well consistent with the results for all the 9897 
Seyfert-2 galaxies. Therefore, there are few effects of spectral S/N on our final results. 

	For the second point, effects of different redshift on the dependence of D{\it n}4000 on $LR$ in Type-2 AGN is 
checked. Among the 9897 Seyfert-2 galaxies (LINERs not considered), correlations between D{\it n}4000 and $LR$ are carefully 
checked for the 4581 Seyfert-2 galaxies with $z>0.1$ (Spearman rank correlation coefficient about -0.68 with $P_{null}<10^{-15}$) 
and the 5316 Seyfert-2 galaxies with $z<0.1$ (Spearman rank correlation coefficient about -0.69 with $P_{null}<10^{-15}$). Here, 
the critical value $z\sim0.1$ is applied only because there are similar numbers of objects with $z<0.1$ and with $z>0.1$. The 
correlations are shown in Fig.~\ref{cz} with the LTS technique determined linear correlations described by
\begin{equation}
\begin{split}
	&D{\it n}4000=(10.088\pm0.104)-(1.052\pm0.013)LR \ \ (z<0.1)\\
	&D{\it n}4000=(10.582\pm0.116)-(1.108\pm0.014)LR \ \ (z>0.1)
\end{split}
\end{equation}
with the RMS scatters about 0.116 and 0.096, respectively, well consistent with the results for all the 9897 Seyfert-2 galaxies. 
Therefore, for Type-2 AGN with redshift less than 0.35, there are no evolution effects on the correlation between D{\it n}4000 
and $LR$. 

\begin{figure*}
	\centering\includegraphics[width = \textwidth]{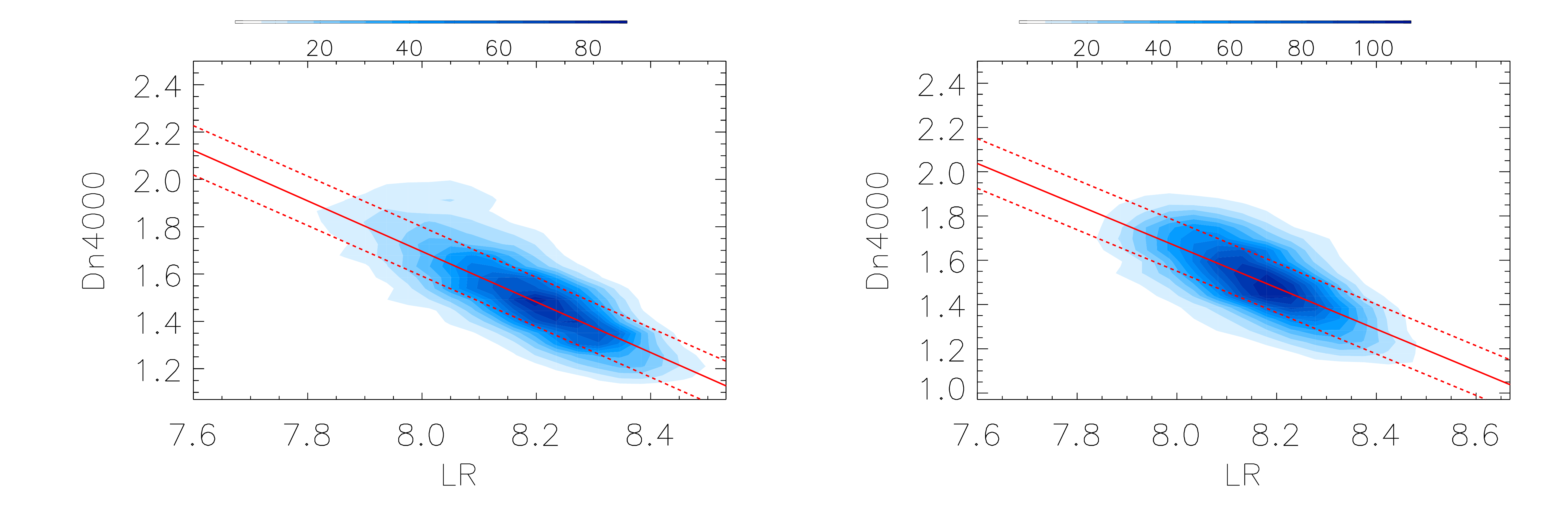}
\caption{On the correlation between D{\it n}4000 and $LR$ in the Seyfert-2 galaxies with $\log(\rm O3HB)<0.56$ 
(left panel) and with $\log(\rm O3HB)>0.56$ (right panel). In each panel, solid red line, dotted red lines show the best 
fitting results to the correlation and the corresponding 1RMS scatters, respectively.}
\label{co3hb}
\end{figure*}

\begin{figure*}
	\centering\includegraphics[width = \textwidth]{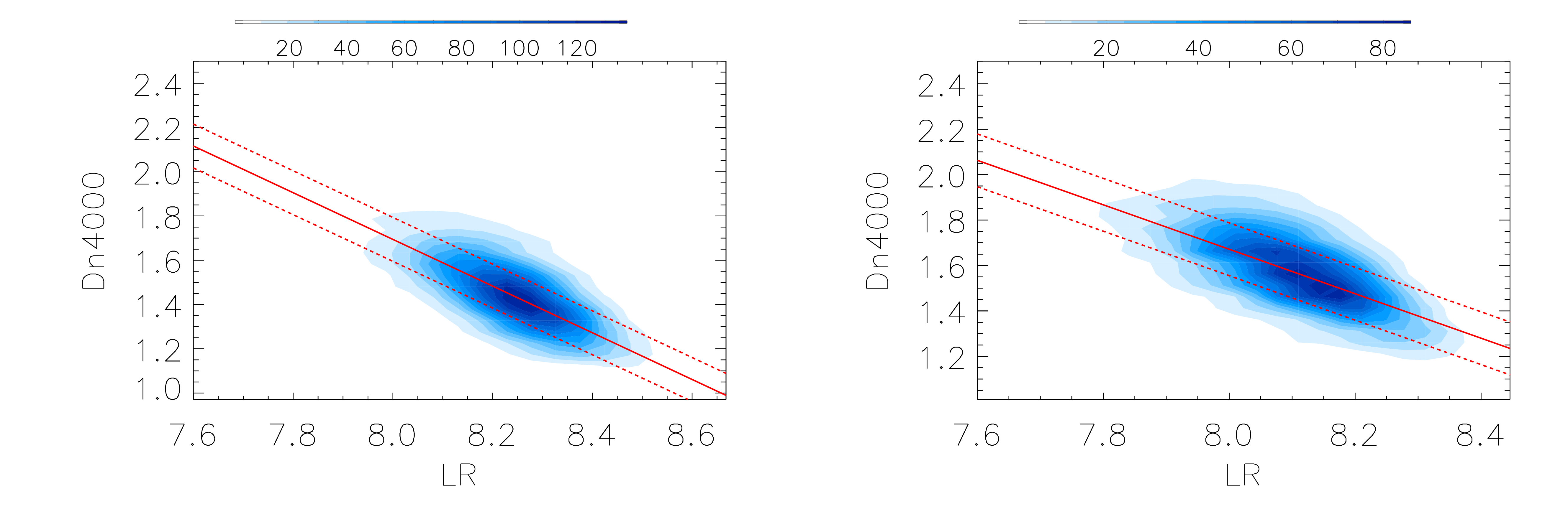}
\caption{On the correlation between D{\it n}4000 and $LR$ in the Seyfert-2 galaxies with $\log(\rm N2HA)<-0.091$ 
(left panel) and with $\log(\rm N2HA)>-0.091$ (right panel). In each panel, solid red line, dotted red lines show the 
best fitting results to the correlation and the corresponding 1RMS scatters, respectively.}
\label{cn2ha}
\end{figure*}

	For the third point, dependence of the correlation between D{\it n}4000 and $LR$ in Type-2 AGN on central activity 
traced by O3HB is checked. Among the 9897 Seyfert-2 galaxies (LINERs not considered), correlations between D{\it n}4000 
and $LR$ are carefully checked for the 4925 Seyfert-2 galaxies with $\log(\rm O3HB)<0.56$ (Spearman rank correlation 
coefficient about -0.77 with $P_{null}<10^{-15}$) and the 4972 Seyfert-2 galaxies with $\log(\rm O3HB)>0.56$ (Spearman 
rank correlation coefficient about -0.68 with $P_{null}<10^{-15}$). Here, the critical value $\log(\rm O3HB)\sim0.56$ is 
applied only because there are similar numbers of objects with $\log(\rm O3HB)<0.56$ and with $\log(\rm O3HB)>0.56$. 
The correlations are shown in Fig.~\ref{co3hb} with the LTS technique determined linear correlations described by
\begin{equation}
\begin{split}
	&D{\it n}4000~=~(10.239\pm0.093)~-~(1.068\pm0.012)LR \hspace{.5cm} (\log(\rm O3HB)<0.56)\\
	&D{\it n}4000~=~(9.148\pm0.099)~-~(0.936\pm0.013)LR \hspace{.5cm} (\log(\rm O3HB)>0.56)
\end{split}
\end{equation}
with the RMS scatters about 0.104 and 0.112, respectively, well consistent with the results for all the 9897 Seyfert-2 galaxies. 
Therefore, for the Type-2 AGN in the BPT diagram, there are no dependence of the correlation between D{\it n}4000 and 
$LR$ on central activity.

\begin{figure*}
	\centering\includegraphics[width = \textwidth]{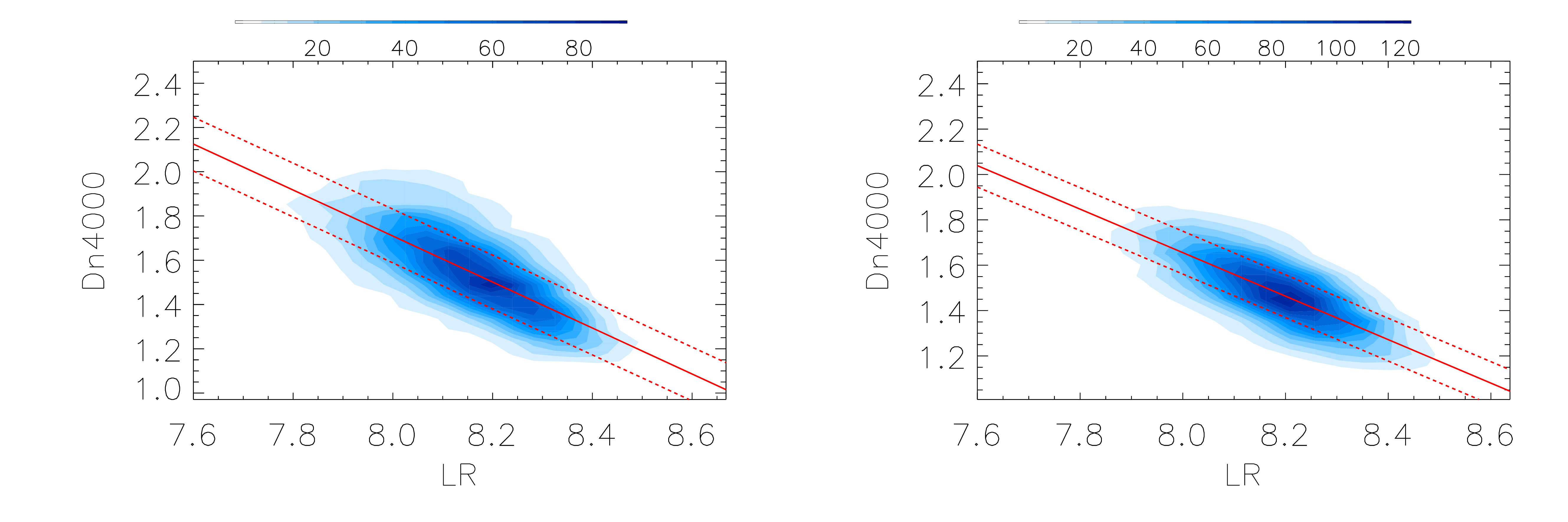}
\caption{On the correlation between D{\it n}4000 and $LR$ in the Seyfert-2 galaxies with $IC<0.374$ (left panel) 
and with $IC>0.374$ (right panel). In each panel, solid red line, dotted red lines show the best fitting results to the 
correlation and the corresponding 1RMS scatters, respectively.}
\label{cmor}
\end{figure*}

	For the fourth point, it is necessary to check probable effects of the applied range of N2HA on the dependence of 
D{\it n}4000 on $LR$. Among the 9897 Seyfert-2 galaxies (LINERs not considered), correlations between D{\it n}4000 
and $LR$ are carefully checked for the 4987 Seyfert-2 galaxies with $\log(\rm N2HA)<-0.091$ (Spearman rank correlation 
coefficient about -0.72 with $P_{null}<10^{-15}$) and the 4910 Seyfert-2 galaxies with $\log(\rm N2HA)>-0.091$ (Spearman 
rank correlation coefficient about -0.62 with $P_{null}<10^{-15}$). Here, the critical value $\log(\rm N2HA)\sim-0.091$ 
is applied only because there are similar numbers of objects with $\log(\rm N2HA)<-0.091$ and with 
$\log(\rm N2HA)>-0.091$. The correlations are shown in Fig.~\ref{cn2ha} with the LTS technique determined linear 
correlations described by
\begin{equation}
\begin{split}
	&D{\it n}4000~=~(10.133\pm0.099)~-~(1.055\pm0.012)LR \hspace{.5cm} (\log(\rm N2HA)<-0.091)\\
	&D{\it n}4000~=~(9.502\pm0.121)~-~(0.979\pm0.015)LR \hspace{.5cm} (\log(\rm N2HA)>-0.091)
\end{split}
\end{equation}
with the RMS scatters about 0.099 and 0.116, respectively, well consistent with the results for all the 9897 Seyfert-2 galaxies.
Therefore, for the Type-2 AGN in the BPT diagram, there are notiny effects of the applied range of N2HA on the dependence of 
D{\it n}4000 on$LR$.

	For the fifth point, dependence of the correlation between D{\it n}4000 and $LR$ in Type-2 AGN on host galaxy 
morphology properties traced by inverse concentration parameter $IC=R_{50}/R_{90}$ is checked, where $R_{50}$ and $R_{90}$ represent 
the radii containing 50\% and 90\% of the Petrosian flux in SDSS r band. Among the 9897 Seyfert-2 galaxies (LINERs not considered), 
there are 9853 Seyfert-2 galaxies with reliable SDSS pipeline provided parameters of $R_{50}$ and $R_{90}$ (parameter name of 
'petroR50\_r' and 'petroR50\_r' in in the SDSS database of 'PHOTOOBJALL'). Among the 9853 Seyfert-2 galaxies, correlations 
between D{\it n}4000 and $LR$ are carefully checked for the 4943 Seyfert-2 galaxies with $IC<0.374$ (Spearman rank correlation 
coefficient about -0.71 with $P_{null}<10^{-15}$) and the 4910 Seyfert-2 galaxies with $IC>0.374$ (Spearman rank correlation 
coefficient about -0.73 $P_{null}<10^{-15}$). Here, the critical value $IC\sim0.374$ is applied only because there are similar 
numbers of objects with $IC<0.374$ and with $IC>0.374$. The correlations are shown in Fig.~\ref{cmor} with the LTS technique 
determined linear correlations described by
\begin{equation}
\begin{split}
	&D{\it n}4000=(10.022\pm0.105)-(1.039\pm0.013)LR  \hspace{1.5cm}(IC<0.374)\\
	&D{\it n}4000=(9.326\pm0.089)-(0.959\pm0.011)LR  \hspace{1.5cm}(IC>0.374)
\end{split}
\end{equation}
with the RMS scatters about 0.122 and 0.095, respectively, well consistent with the results for all the 9897 Seyfert-2 galaxies. 
Therefore, for the Type-2 AGN, there are no dependence of the correlation between D{\it n}4000 and $LR$ on host galaxy 
morphology properties.

\begin{figure*}
	\centering\includegraphics[width = \textwidth]{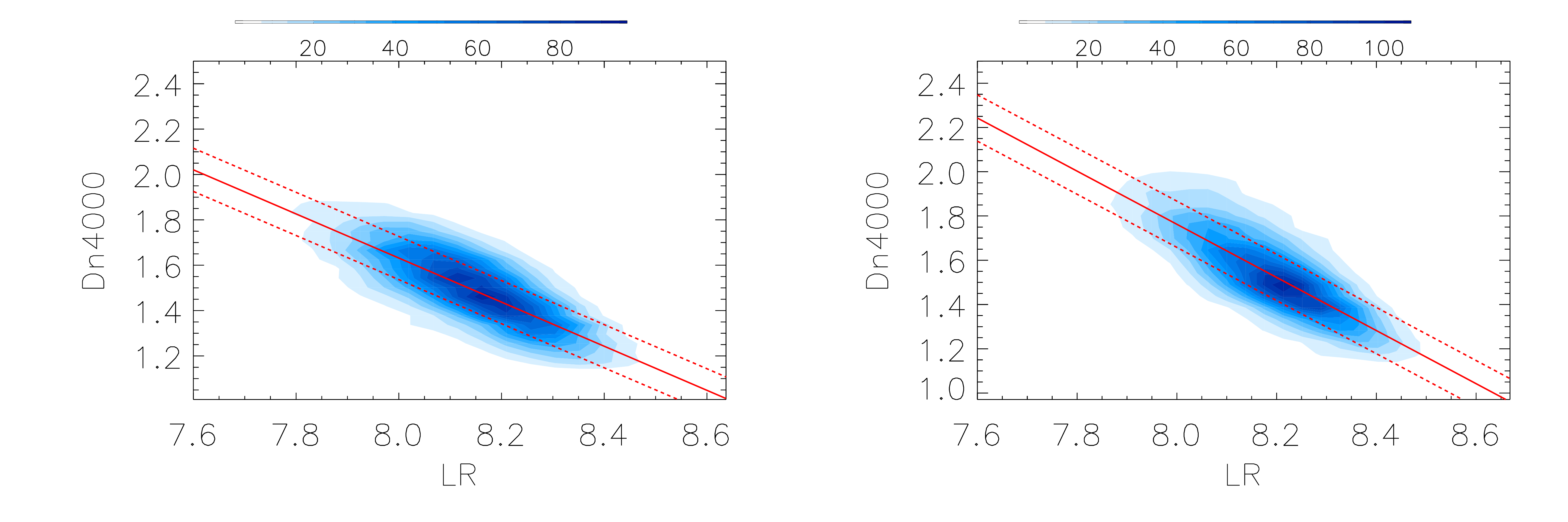}
\caption{On the correlation between D{\it n}4000 and $LR$ in the Seyfert-2 galaxies with $\sigma_\star<135{\rm km/s}$ (left 
panel) and with $\sigma_\star>135{\rm km/s}$ (right panel). In each panel, solid red line, dotted red lines show the best fitting 
results to the correlation and the corresponding 1RMS scatters, respectively.}
\label{cste}
\end{figure*}

	For the sixth point, dependence of the correlation between D{\it n}4000 and $LR$ in Type-2 AGN on stellar velocity 
dispersion $\sigma_\star$ is checked. Considering the tight connection between stellar velocity dispersion and central BH mass 
(the known M-sigma relation) \citep{fm00, ge00, mm13, ref3} and the tight connection between total stellar mass and central BH mass 
\citep{mh03, hr04, mm13, ref3, sh20}, the dependence on stellar velocity dispersion will provide clues on effects of the BH mass 
and total stellar mass on the correlation between D{\it n}4000 and $LR$ in Type-2 AGN. Among the 9897 Seyfert-2 galaxies 
(LINERs not considered), correlations between D{\it n}4000 and $LR$ are carefully checked for the 4877 Seyfert-2 galaxies with 
$\sigma_\star<135{\rm km/s}$ (Spearman rank correlation coefficient about -0.76 with $P_{null}<10^{-15}$) and the 5020 Seyfert-2 
galaxies with $\sigma_\star>135{\rm km/s}$ (Spearman rank correlation coefficient about -0.77 with $P_{null}<10^{-15}$). Here, 
the critical value $\sigma_\star\sim135{\rm km/s}$ is applied only because there are similar numbers of objects with 
$\sigma_\star<135{\rm km/s}$ and with $\sigma_\star>135{\rm km/s}$. The correlations are shown in Fig.~\ref{cste} with the LTS 
technique determined linear correlations
\begin{equation}
\begin{split}
	&D{\it n}4000~=~(9.408\pm0.084)~-~(1.006\pm0.011)LR \hspace{.5cm} (\sigma_\star<135{\rm km/s}) \\
	&D{\it n}4000~=~(11.365\pm0.096)~-~(1.201\pm0.012)LR \hspace{.5cm} (\sigma_\star>135{\rm km/s})
\end{split}
\end{equation}
with the RMS scatters about 0.095 and 0.105, respectively, consistent with the results for all the 9897 Seyfert-2 galaxies. 
Therefore, for the Type-2 AGN, there are no dependence of the correlation between D{\it n}4000 and $LR$ on host galaxy 
stellar velocity dispersions.

\begin{figure*}
	\centering\includegraphics[width = \textwidth]{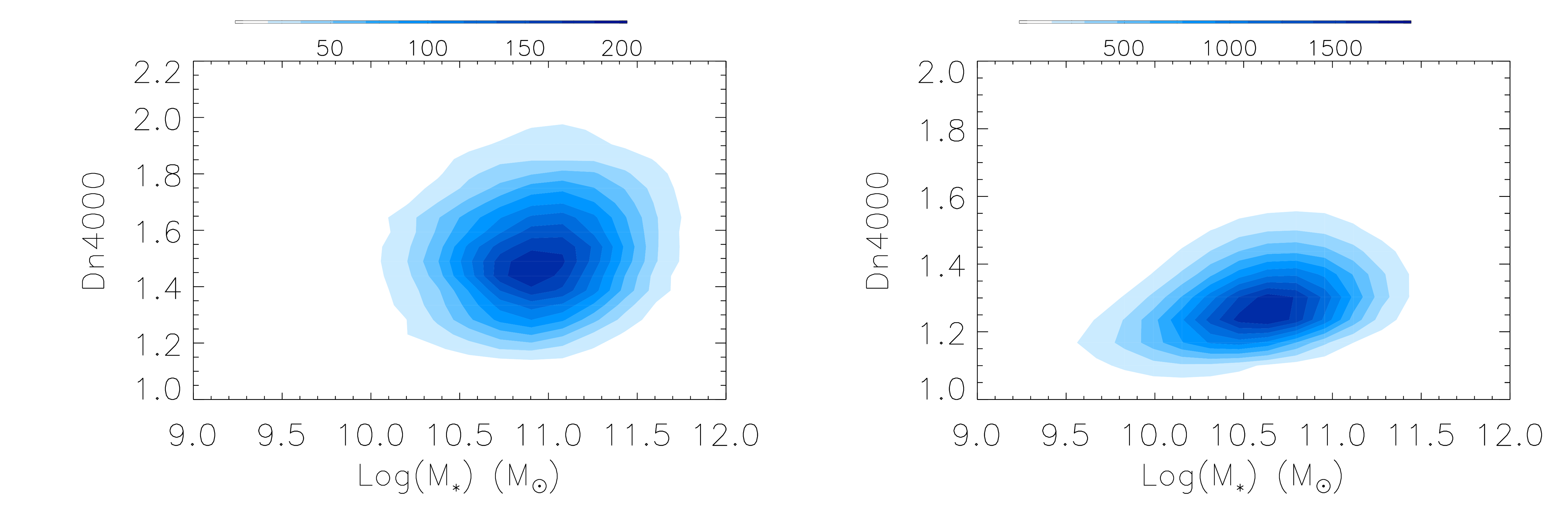}
\caption{On the correlation between D{\it n}4000 and total stellar mass $M_\star$ in the Seyfert-2 galaxies (left panel) and in 
the HII galaxies (right panel).}
\label{mass}
\end{figure*}

	For the seventh point, motivated by the reported dependence of the D{\it n}4000 on the total stellar mass in 
\citet{ref21, zg17}, it is interesting to check which parameter, the parameter $LR$ or the total stellar mass $M_\star$, is 
preferred to trace the D{\it n}4000. Here, the public database 'StellarMassPCAWiscBC03' \citep{ch12} from MPA/JHU research 
group is accepted to collect the measured total stellar masses of narrow emission line galaxies in SDSS. Then, Fig.~\ref{mass} 
shows dependence of D{\it n}4000 on the total stellar mass of the collected 9897 Seyfert-2 galaxies (LINERs not considered) 
and of the collected 44501 HII galaxies, with Spearman rank correlation coefficients about 0.15 ($P_{null}<10^{-15}$) and 0.39 
($P_{null}<10^{-15}$) respectively. The results are simply consistent with the results in \citet{ref21} that strong AGN (here, 
Seyfert-2 galaxies) have weak dependence of the D{\it n}4000 on the $M_\star$. Due to the weakness of the dependence in 
Fig.~\ref{mass}, there are no descriptions on the dependence by formula. The different dependence of D{\it n}4000 on 
$M_\star$ in Seyfert-2 galaxies and in HII galaxies could be due to AGN feedback leading host galaxy evolution of AGN different 
from the normal evolution of quiescent galaxies. To discuss the different dependence of D{\it n}4000 on $M_\star$ by 
effects of AGN activity is not the objective of this manuscript. However, the results can be applied to support that total 
stellar mass can be simply applied to trace stellar age (D{\it n}4000) only in quiescent galaxies, whereas the parameter 
$LR$ can be efficiently applied to trace stellar age (D{\it n}4000) in both AGN and none-AGN galaxies.


\section{Summary and Conclusions}

	The final summary and conclusions are as follows. 
\begin{itemize}	
\item Based on well measured narrow emission lines of low redshift narrow emission line main galaxies in SDSS DR16, large samples 
	of 14031 Type-2 AGN and 44501 HII galaxies are collected through applications of BPT diagram of O3HB versus N2HA. 
\item Strong positive correlation between D{\it n}4000 and N2HA and negative correlation between D{\it n}4000 and 
	O3HB can be confirmed in the Type-2 AGN. However, weaker corresponding correlations between D{\it n}4000 and narrow 
	emission line ratios are detected in HII galaxies.
\item Strong negative correlations can be confirmed between D{\it n}4000 and narrow emission line luminosities in both Type-2 AGN 
	and HII galaxies.
\item Combining N2HA and narrow H$\alpha$ line luminosity, the stronger negative correlation between D{\it n}4000 and 
	$LR=0.2\log(L_{H\alpha}/{\rm erg\cdot s^{-1}})-0.5\log(\rm N2HA)$ can be confirmed in Type-2 AGN and 
	in HII galaxies, with smaller RMS scatters.
\item Through applications of BPT diagrams, LINERs have weaker dependence of D{\it n}4000 on narrow emission line ratios and 
	narrow line luminosities, indicating different intrinsic physical properties of LINERs from Seyfert-2 galaxies.
\item There are consistent correlations between D{\it n}4000 and $LR$ for the Type-2 AGN with different spectral S/N, different 
	redshifts, different O3HB, different N2HA, different host galaxy morphology properties and different host galaxy stellar 
	velocity dispersions, leading to the robust and strong dependence of D{\it n}4000 on the parameter of $LR$ in Type-2 AGN.
\item Applications of narrow emission line properties to trace the parameter D{\it n}4000 in Type-2 AGN will provide a convenient 
	method to estimate statistical properties of stellar ages of samples of more luminous AGN with weak host galaxy 
	absorption features but with apparent narrow emission lines.
\end{itemize}

\begin{acknowledgements}
Zhang gratefully acknowledge the anonymous referee for giving us constructive comments and suggestions to greatly improve the paper. 
Zhang gratefully thanks the kind financial support from GuangXi University and the kind grant support from NSFC-12173020 and 
NSFC-12373014 and the the Guangxi Talent Programme (Highland of Innovation Talents). This manuscript has made use of the data 
from the SDSS projects. The SDSS-III web site is http://www.sdss3.org/. SDSS-III is managed by the Astrophysical Research Consortium 
for the Participating Institutions of the SDSS-III Collaborations.
\end{acknowledgements}

\end{document}